\documentclass[aps,pre,epsfigure,superscriptaddress,nofootinbib,notitlepage]{revtex4-1}
\usepackage[colorlinks=true,linkcolor=blue,urlcolor=blue,citecolor=blue,pdfusetitle]{hyperref}
\usepackage[utf8]{inputenc}
\usepackage[english]{babel}
\usepackage{amsmath}
\usepackage[caption = false]{subfig}
\usepackage{verbatim}
\usepackage{graphicx,epstopdf}
\usepackage{blindtext}
\usepackage[table,xcdraw]{xcolor}
\usepackage{lipsum}
\usepackage{amsfonts}
\usepackage{bbm}
\usepackage{amssymb}
\usepackage{enumerate}
\usepackage{color}
\usepackage{latexsym}
\usepackage{physics}
\usepackage{cancel}
\usepackage{soul}
\usepackage{times,txfonts}
\usepackage{braket}
\usepackage{soul}
\usepackage{natbib}

\newcommand{\1}{\mathbbm{1}}

\usepackage{setspace} 

\begin{document}

\doublespacing  

\title{Quantum thermodynamics of a power-law potential}

\author{Vinicius Gomes de Paula}
\email{viniciusgomespaula@id.uff.br}
\affiliation{Physics Institute, Fluminense Federal University, Av. Gal. Milton Tavares de Souza s/n, 24210-346 Niter\'{o}i, Rio de Janeiro, Brazil.}

\author{Wanisson S. Santana}
\email{wanisson.santana@ufob.edu.br}
\affiliation{Physics Institute, Fluminense Federal University, Av. Gal. Milton Tavares de Souza s/n, 24210-346 Niter\'{o}i, Rio de Janeiro, Brazil.}
\affiliation{Quantum Information and Statistical Physics Group, Western Bahia Federal University - Campus Reitor Edgard Santos. Rua Bertioga, 892, Morada Nobre I, 47810-059 Barreiras, Bahia, Brasil.}

\author{Clebson Cruz}
\email{clebson.cruz@ufob.edu.br}
\affiliation{Quantum Information and Statistical Physics Group, Western Bahia Federal University - Campus Reitor Edgard Santos. Rua Bertioga, 892, Morada Nobre I, 47810-059 Barreiras, Bahia, Brasil.}

\author{Mario Reis}
\email{marioreis@id.uff.br}
\affiliation{Physics Institute, Fluminense Federal University, Av. Gal. Milton Tavares de Souza s/n, 24210-346 Niter\'{o}i, Rio de Janeiro, Brazil.}

\begin{abstract}
\setstretch{1.3}
Modeling quantum thermal machines provides a practical approach to describing the thermodynamic properties of quantum technologies and devices. For this purpose, 
power-law potentials are often employed as working mediums of quantum thermodynamic cycles to investigate the concepts of heat, work, and efficiency. With this in mind, we present the results for the Stirling and Otto numerical modeling of quantum thermal machines that use a general power law potential with a characteristic $q$ exponent. 
We calculate its energy spectra, showing that it recovers the traditional forms of harmonic oscillator 
and 1-D potential well. 
We derive expressions for the reduced energy exchanges during a complete cycle and for the efficiency/coefficient of performance as a function of the exponent $q$, the bath temperatures, and the frequency ratio. From these results, we identify parameters that yield desired properties, such as optimized performance and transitions between different operation modes. The findings highlight the role of power-like potentials in optimizing quantum heat engines and support the design of tailored engines with specific performance characteristics.

\end{abstract}

\maketitle

\section{Introduction}

The growing demand for efficient energy transfer and storage techniques in quantum devices is a modern technological challenge \cite{metzler2023emergence,PRXQuantum.3.020101}. In this scenario, the study of quantum cycles offers valuable insights into the behavior of thermodynamic systems at the nanoscale, where quantum effects dominate and classical mechanics may not apply \cite{myers2022quantum}. Understanding quantum cycles can also shed light on the development of more efficient energy conversion technologies and advanced quantum systems \cite{cruz2022quantum,cruz2023quantum,campaioli2023colloquium,metzler2023emergence,PRXQuantum.3.020101,myers2022quantum}. As a consequence, quantum thermodynamics emerged due to technological advancements enabling the miniaturization of devices and materials \cite{myers2022quantum,Gemmer:Book,Deffner-Campbel:Book}. Particularly, it is a powerful theoretical tool to address work extraction of small-scale quantum thermal machines based on real materials \cite{cruz2022quantum,cruz2023quantum}. Understanding the roles played by both heat and work on the quantum scale is crucial for this. Leveraging the quantum properties of the working medium can lead to many extraordinary results with practical applications in energy harvesting and quantum information \cite{bernardo2020unraveling, myers2022quantum}.

For this task, the modeling of quantum cycles using power-like potentials  \cite{mahajan2020quantum, aydiner2021quantum, islam2023many} can provide an insightful framework to investigate the physical concepts of thermodynamics, such as heat, work and efficiency, on the quantum scale. This is because many physically meaningful potentials can be described by a polynomial function or can be expanded in terms of it. Therefore, utilizing  power-like potentials in quantum cycle modeling can offer a versatile approach to studying thermodynamic principles in different quantum systems. These models can help researchers gain a deeper understanding of fundamental principles governing energy transfer and conversion  in quantum systems, exploring complex interactions between particles and fields, and shedding light on the behavior of quantum systems under various conditions. By investigating the impact of different parameters on the efficiency and performance of quantum systems, these models can lead to potential advancements in quantum technology and energy conversion processes \cite{myers2022quantum}. 

In this context, we propose numerical modeling of Stirling and Otto quantum cycles for a power-like potential of the $V(x)=V_0\left(\frac{x}{2a}\right)^{2q}$ form, where $q=1,2,3..$, $V_0$ is the potential strength and $a$ is the spatial scale for the generalized well. We refer to it as a generalized potential, since it is a generalization of both the quantum harmonic oscillator ($q=1$) and the 1-D infinite square well potential ($q\rightarrow \infty$). Thus, any other 1-D potential can be recovered by imposing the appropriate limits on the $q$ values in this potential. In this regard, we investigated the role played by the exponent $q$ in thermodynamic behavior of our generic power-like potential in the search of physical potentials with desirable properties while operating as working substance under Stirling and Otto quantum cycles. The thermodynamic quantities display a strong dependence on the $q$ parameter up to $q=3$ value, above which the behavior of the different potentials are dictated by the other physical parameters. For the Stirling cycle we observed that adjusting the values of $T_{h}$ and $T_{c}$ the system can alternate between heater and refrigerator modes. On the other hand, for the Otto cycle the tuning of the oscillating frequency ratio ($\omega_h/\omega_c$) allows maximizes the work output and alternate between a heat engine and refrigerator modes.  Thus, the results shown derived mathematical expressions for energy exchanges and efficiency as functions of the temperatures of the thermal reservoirs and the parameter $q$. The findings provide valuable insights into the optimization of quantum heat engines, highlighting the role of power-law potentials in improving the performance characteristics of quantum thermal machines. From a physical point of view, this approach provides a systematic way to search for physical systems with improved quantum thermodynamic properties.

\section{Energy spectra of working medium }\label{sec:WorkingMedium}

The working substance is represented by a potential of the following form:
\begin{align} \label{gen_pot_x2q}
  V(x)=V_0\left(\frac{x}{a}\right)^{2q}  
\end{align}
where $q = 1,2,3...$; $V_0$ is a constant parameter that sets the overall scale of the potential, determining the strength (or height) of the potential function; and $a$ is a constant that provides the spatial scale to the potential. This potential contains two well-studied cases, the harmonic oscillator ($q=1$) \cite{chatterjee2021temperature} and the particle in a box ($q \rightarrow \infty$) \cite{thomas2019quantum}. Here, fixing the $q$ value defines a specific potential. Therefore, explicitly writing the potential as a function of the $q$ parameter allows us to map the behavior of the thermodynamic quantities of interest for different power-like potentials.

\begin{figure}[ht]
		\centering
		\includegraphics[scale=0.3]{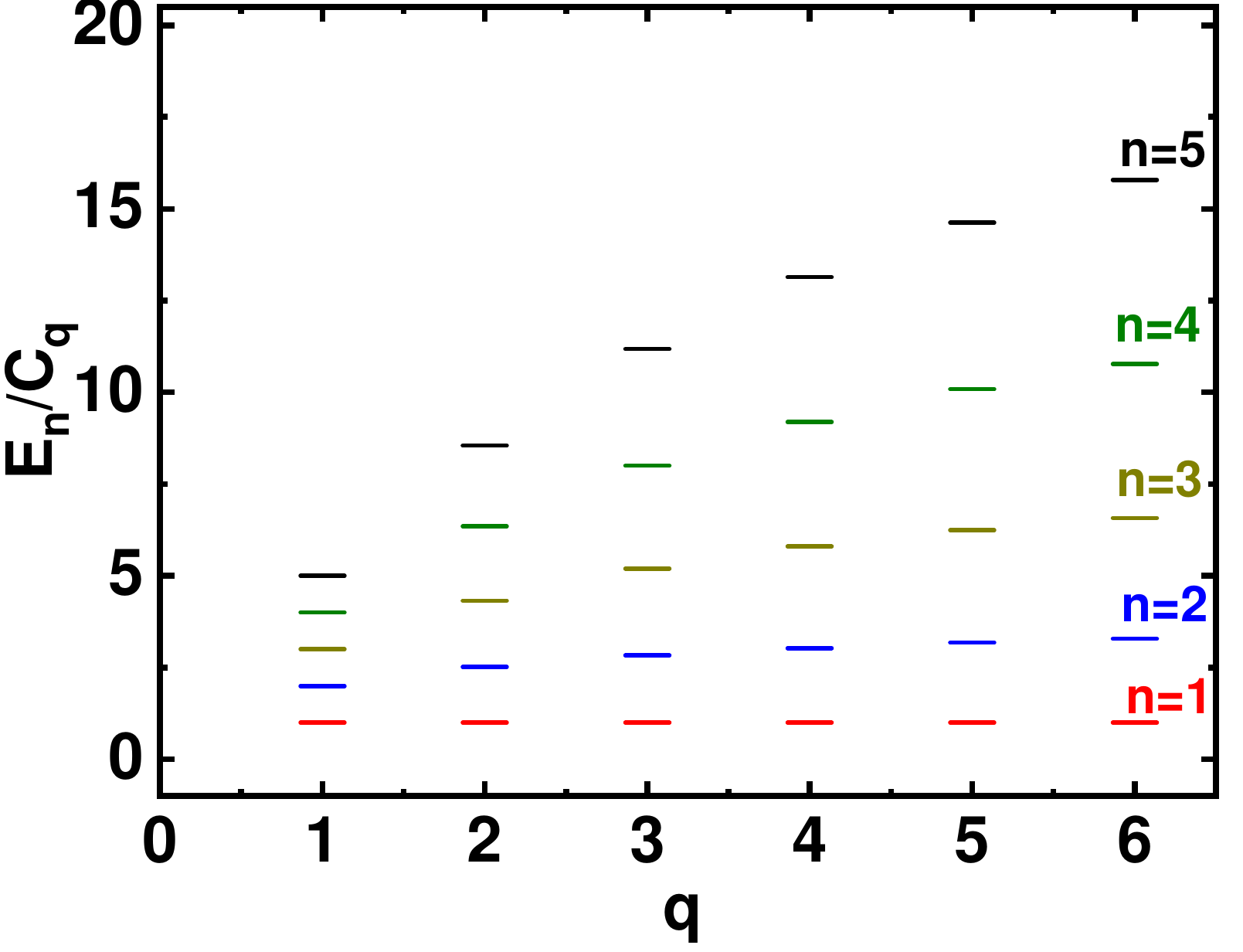}
		\caption{Energy levels for integer values of $n$ calculated as function of the $q$ parameter, calculated from Equation \ref{En}, showing the increase in the energy gap as the $q$ value increases from linear to quadratic. }
    \label{Figure1_Enlevels}
\end{figure}

Semi-classical quantization rules give us the following expression for the energy levels \cite{mario2025}:
\begin{align} \label{En}
    E_n=n^{\frac{2q}{1+q}}C_q
\end{align}
where the constant $C_q$ is given as:
\begin{align} \label{Cq}
    C_q=\left(\frac{ \hbar}{2ma^2} \right)^{\frac{q}{q+1}}V{_0}^{\frac{1}{1+q}} \left(\frac{\sqrt{\pi}\; \Gamma(3/2 + 1/2q)}{ \Gamma(1 + 1/2q)}\right)^{\frac{2q}{1+q}}
\end{align}
Using the energy spectra given by equation \ref{En}, we can plot the energy levels for different $n$ values as shown in the Figure \ref{Figure1_Enlevels}.
We can see that the separation among energy levels (i.e. the energy gaps) increases as we increase the $n$ value. From equation \ref{En}, we see that the exponent of $n$ is given by the term $2q/(q+1)$, therefore the gap of energy levels can be tuned from linear to quadratic, as we go from $q=1$ up to $q \rightarrow \infty$. It is straightforward to show that:
\begin{align} \label{q1limit}
    E_n(q=1) =n \hbar \omega
\end{align}
where $\omega=\frac{2V_0}{ma^{2}}$ is the frequency of the quantum harmonic oscillator. Moreover, we also have:
\begin{align} \label{qifnlimit}
   \lim_{q\to\infty} E_n(q) = \left( \frac{\pi^2\hbar^2}{2ma^2} \right) \frac{n^{2}}{4}
\end{align}
which are the energy spectra for the harmonic oscillator (linear gap) and a particle in a box (quadratic gap), respectively. The exponent of the parameter $V_0$ is $1/(1+q)$, which varies between $1/2$ (for $q = 2$) and $0$ (for $q \rightarrow \infty$). Therefore, the separation of energy levels with the quantum number $n$ can range from linear to quadratic, depending on the $q$ value. However, the energy spectrum becomes increasingly independent of the parameter $V_0$ as the value of $q$ increases.


We will use the Equation \ref{En} to obtain the thermodynamic quantities required to model both Stirling and Otto quantum cycles. The first step it to write a expression for the partition function for a quantum thermal state at temperature $T$:
\begin{align} \label{Z_thermal_states}
    Z(q,T)=\sum_{n=1}^{\infty} \exp\left(-\frac{E_n}{k_B
 T}\right)=\sum_{n=1}^{\infty} \exp\left({-\frac{n^{\frac{2q}{1+q}}C_q}{k_B T}}\right) = \frac{1}{2} \Theta\left[0,\exp\left(-\frac{n^{\frac{2q}{1+q}}C_q}{k_{B} T}\right)\right]
\end{align}
where the summation over the $n$ states is written in terms of a theta Jacobi function $\Theta$ defined as: 
\begin{align} \label{jacobi_function}
    \sum_{m=1}^{\infty} \exp(-\alpha m^2) = \frac{1}{2} \Theta\left[0,\exp(-\alpha m^2)\right].
\end{align}
Complementarily, the probability occupation for the $n^{th}$ thermal state is given by the following expression:
\begin{align} \label{pn_thermal_states}
    p_n=\frac{\exp\left(-\frac{E_n}{k_B T}\right)}{Z(q,T)}=\frac{\exp\left({-\frac{n^{\frac{2q}{1+q}}C_q}{k_B T}}\right)}{Z(q,T)}
\end{align}
The energy spectra $E_n$, the partition function $Z(q,T)$, and the probability occupation $p_n$ are the building blocks to model quantum thermodynamic cycles. We will use them to obtain mathematical expressions for the heat changes, net work per cycle, and efficiency of both Stirling and Otto quantum cycles, as a function of the $q$ parameters, and the thermal baths $T_{h}$ and $T_{c}$.

\section{Quantum Stirling cycle} \label{sec:StirlingCycle}

The classical Stirling cycle is composed of two isothermal and two isochoric processes that are used to represent a reversible heat transfer device, displaying the Carnot efficiency. The quantum version of the cycle, like its classical counterpart, is also a four-stroke closed-cycle regenerative heat engine, consisting of two (quantum) isothermal and two (quantum) isochoric processes. One main advantage of the quantum Stirling cycle, compared to other quantum cycles, is that it does not involve a quantum adiabatic stroke, which is particularly challenging to implement experimentally.

Quantum Stirling heat machines have already been theoretically explored, using various working mediums such as coupled spin systems \cite{huang2014quantum, rojas2024quantum, pili2023quantum}, sets of harmonic oscillators \cite{evkaya2023quantum,papadatos2023quantum,rodin2024three}, and potential well \cite{sur2023quantum,das2023quantum}. Among the conclusions obtained in these earlier studies is that Stirling quantum thermal machines can be settled to operate under different regimes such as a regenerator or heat engine, a feature not observed in the classical Stirling cycle.

For modeling the quantum Stirling cycle for the generalized power-law potential (Equation \ref{gen_pot_x2q}), we will closely follow the approach described in references \cite{chatterjee2021temperature} and \cite{thomas2019quantum}, whose leverages some symmetry properties of the potential to calculate the relevant thermodynamic properties. In this approach, the quantum isothermal expansion (compression) process is taken as an insertion (removal) of a barrier in the middle of the potential well under quasi-static conditions. The net effect is that the energy levels associated with even values of $n$ remain unchanged while each energy level with an odd value of $n$ shifts upwards and overlaps with the nearest neighboring even energy level of the original single well potential. This leads to a degeneracy in the energy levels of this new double well setup and a multiplicity of the partition functions. In this approach, it is convenient to define the following auxiliary partition functions:
\begin{align} \label{Z_h}
Z_{T_h}=\sum_{n=1}^{\infty} \exp\left({-\frac{n^{\frac{2q}{1+q}}C_q}{k_B T_h}}\right) = \frac{1}{2} \Theta\left[0,\exp\left(-\frac{n^{\frac{2q}{1+q}}C_q}{k_{B} T_h}\right)\right] = \frac{1}{2} \Theta_{h}
\end{align}
and
\begin{align} \label{Z_c}
Z_{T_c}=\sum_{n=1}^{\infty} \exp\left({-\frac{n^{\frac{2q}{1+q}}C_q}{k_B T_c}}\right) = \frac{1}{2} \Theta\left[0,\exp\left(-\frac{n^{\frac{2q}{1+q}}C_q}{k_{B} T_c}\right)\right] = \frac{1}{2}\Theta_{c}
\end{align}
which correspond to thermal states in equilibrium with the hot and cold bath, respectively. Then, we can write the following partition functions in terms of $Z_{T_h}$ and $Z_{T_c}$ functions (see details in Appendix \ref{AP:Stirling}):
\begin{align} \label{Z_A_Stirling}
Z_A = \exp\left(\frac{-4q}{q+1}\right) Z_{T_h}
\end{align}
\begin{align} \label{Z_B_Stirling}
Z_B = 2 Z_{T_h}
\end{align}
\begin{align} \label{Z_C_Stirling}
Z_C = 2 Z_{T_c}
\end{align}
\begin{align} \label{Z_D_Stirling}
Z_D = \exp\left(\frac{-4q}{q+1}\right) Z_{T_c}
\end{align}
where A, B, C,and D correspond to the four thermalized states of the Stirling cycle (see details in the next subsections \ref{sec:AB_Stirling}  throughout \ref{sec:DA_Stirling}). Writing the partition functions in this form facilitates the calculations of the thermodynamic quantities for each step of the quantum Stirling cycle (see Appendix \ref{AP:Stirling}).

\subsection{Isothermal expansion} \label{sec:AB_Stirling}

Initially, the working substance is in thermal equilibrium with the hot thermal bath at temperature $T_{h}$, while it experiences a quasi-static insertion of an infinite potential barrier at the center. This isothermal process introduces degenerescence in the energy spectra of the system, changing both its energy levels and the corresponding populations. Applying the First Law of thermodynamics gives us:
\begin{align} \label{Q_AB_Stirling}
   \frac{\Delta Q_{AB}(q,T_h,T_c)}{k_{B} T_{c}}= -\frac{T_{h}}{T_{c}} \left[\ln(2) + \frac{4q}{q+1} \right]
\end{align}
for the reduced heat change that occurs in the A $\rightarrow$ B isotherm process. It is important to keep in mind that Equation \ref{Q_AB_Stirling} accounts for the changes in both internal energy and work of the system. During this process, both $\Delta Q_{A,B}$ and $\Delta W_{A,B}$ change in a way that $\Delta U_{A,B}$ does not necessarily remain constant.

\subsection{Isochoric cooling} \label{sec:BC_Stirling}

The system is initially in equilibrium with the hot thermal bath at ${T_h}$. The second step is an isochoric process where no work is done, and the 1º first law gives us:
\begin{align} \label{Q_BC_Stirling}
   \frac{\Delta Q_{BC}(q,T_h,T_c)}{k_{B} T_{c}} = \frac{ C_q}{2 k_{B} T_{c}} \sum_{n=1}^{\infty} n^{\frac{2q}{q+1}}\left\{\frac{\exp\left(-\frac{ C_qn^{\frac{2q}{q+1}}}{k_B T_c}\right)}{ \Theta_{c}} + \frac{ \exp\left(-\frac{ C_qn^{\frac{2q}{q+1}}}{k_B T_h}\right)}{ \Theta_h} \right\}
\end{align}
After the isochoric stroke, the system reaches thermal equilibrium with the cold thermal bath at a temperature $T_{c}<T_{h}$. Thus, as its classical counterpart in isochoric cooling, no work is done, and the system loses heat, calculated by Equation \ref{Q_BC_Stirling}.

\subsection{Isothermal compression} \label{sec:CD_Stirling}
The third step is isothermal, corresponding to the quasi-static removal of the infinite barrier. The system is initially in thermal equilibrium with the cold thermal bath at $T_{c}$. Therefore, according to the 1° law of thermodynamics, we can write:
\begin{align} \label{Q_CD_Stirling}
  \frac{\Delta Q_{CD}(q)}{k_{B} T_{c}} = -\frac{4q}{q+1}
\end{align}
for the reduced heat change between the system and its surroundings for this step. We draw attention to the fact that this quantity is independent of the temperatures of both the hot and the cold reservoirs.

\subsection{Isochoric heating}  \label{sec:DA_Stirling}

Finally, the last step is an isochoric process that leads the system to its initial state, at thermal equilibrium with the hot thermal bath at $T_{h}$:
\begin{align} \label{Q_DA_Stirling}
    \frac{\Delta Q_{DA}(q,T_h,T_c)}{k_{B} T_{c}} = \frac{ C_q}{2 k_{B} T_{c}} \sum_{n=1}^{\infty} n^{\frac{2q}{q+1}}\left\{\frac{\exp\left(-\frac{ C_qn^{\frac{2q}{q+1}}}{k_B T_h}\right)}{ \Theta_h } + \frac{ \exp\left(-\frac{ C_qn^{\frac{2q}{q+1}}}{k_B T_c}\right)}{ \Theta_c} \right\}
\end{align}
Just as in the B$\rightarrow$C process, the heat exchanged is equal to the change in the internal energy of the system.

\subsection{Heat fluxes, work and efficiency for Stirling cycle} \label{Q_in_Q_out_Wnet_eta_Stirling}

For a suitable analysis of the thermodynamic properties of the system, we have also computed the reduced heat fluxes the $Q_{in}/k_{B} T_{c}$ and $Q_{out}/k_{B} T_{c}$, which accounts for the total heat changed with the hot and cold reservoirs, respectively:
\begin{align} \label{Q_in_Stirling}
   \Delta Q_{in}(q,T_h,T_c) =  -\frac{T_{h}}{T_{c}} \left[\ln(2) + \frac{4q}{q+1} \right] +  \frac{ C_q}{2 k_{B} T_{c}} \sum_{n=1}^{\infty} n^{\frac{2q}{q+1}}\left\{\frac{\exp\left(-\frac{ C_qn^{\frac{2q}{q+1}}}{k_B T_h}\right)}{ \Theta_h } + \frac{ \exp\left(-\frac{ C_qn^{\frac{2q}{q+1}}}{k_B T_c}\right)}{ \Theta_c} \right\}
\end{align}

\begin{align} \label{Q_out_Stirling}
    \Delta Q_{out}(q,T_h,T_c) =  \frac{ C_q}{2 k_{B} T_{c}} \sum_{n=1}^{\infty} n^{\frac{2q}{q+1}}\left\{\frac{\exp\left(-\frac{ C_qn^{\frac{2q}{q+1}}}{k_B T_c}\right)}{ \Theta_c } + \frac{ \exp\left(-\frac{ C_qn^{\frac{2q}{q+1}}}{k_B T_h}\right)}{ \Theta_{h}} \right\} -\frac{4q}{q+1} 
\end{align}

The net Work done per cycle is easily calculated by adding the corresponding contribution from both isothermal processes (detailed calculation refer to Appendix \ref{AP:Stirling}):
\begin{align} \label{Wnet_Stirling}
     \frac{\Delta W_{Stir}}{k_{B} T_{c}}  =\frac{T{_h}}{T{_c}}  \ln{\left( \frac {Z_B}{Z_A} \right) } - \ln{\left( \frac {Z_D}{Z_C}\right)}= \frac{T_{h}}{T_{c}} \left[\ln (2) + \frac{4q}{q+1} \right] + \ln (2) + \frac{4q}{q+1}
\end{align}
where the first (second) term is expected to be positive (negative) since it represents Work done by (on) the system. It is the same dependency exhibited by the net Work per cycle for the cases of the harmonic oscillator and the particle in a box.

\begin{figure*}
		\centering
		\includegraphics[scale=0.5]{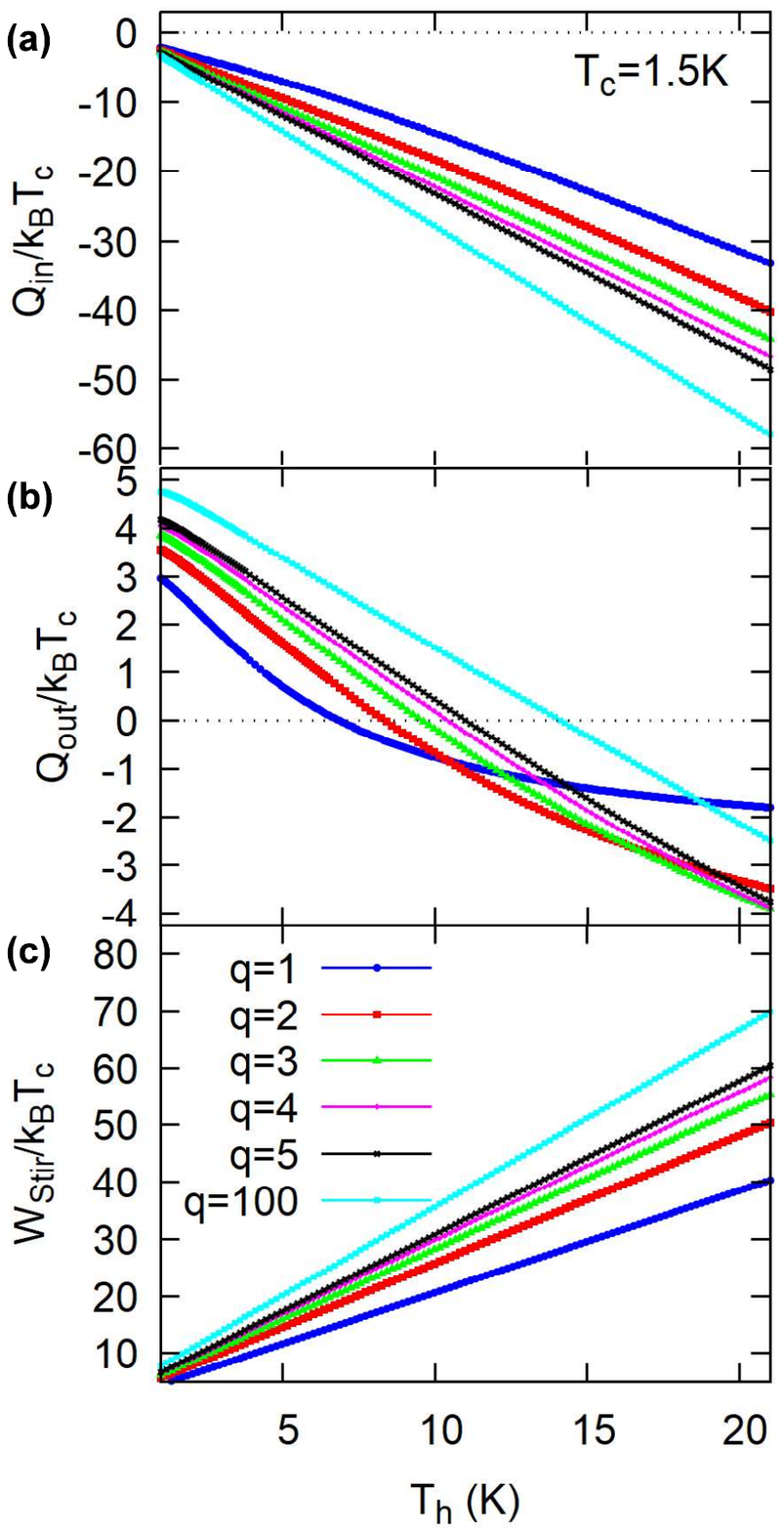}
		\caption{Thermodynamic quantities of interest (a) $Q_{in}/k_{B} T_{c}$, (b) $Q_{out}/k_{B} T_{c}$, and (c) $W_{net}/k_{B} T_{c}$, for the quantum Stirling cycle of the $V(x)=V_0\left(\frac{x}{2a}\right)^{2q}$ potential, plotted as a function of the $T_{h}$ for a fixed cold bath temperature $T_{c}=1.5$ K. Different colors correspond to various $q$ values, each one representing a specific power-law potential. }
    \label{QinQoutWnet_Stir}
\end{figure*}

Investigating the dependence of the reduced quantities $Q_{in}/k_{B} T_{c}$, $Q_{out}/k_{B} T_{c}$, and $W_{Stir}/k_{B} T_{c}$, with the of thermal baths $T_{h}$, and $T_{c}$ allows a complete analysis of the thermodynamic behavior of the system for a fixed $q$ value, i.e. a certain physical potential. For example, the heat flux between the hot bath and the working substance $Q_{in}/k_{B} T_{c}$ decreases as the hot bath $T_h$ increases, for all the $q$ values computed, as we can see from Figure \ref{QinQoutWnet_Stir}(a) considering a cold bath of $T_{c}=1.5$ K. For $q=3$ and higher values, the decrease is linear. Finally, we see that increasing the values of $q$ maximizes the absolute values of $Q_{in}/k_{B} T_{c}$; that is, it increases the heat exchange between the hot bath and the working substance. 

However, the heat flux between the cold bath and the working substance $Q_{out}/k_{B} T_{c}$ shows a distinct behavior as can be visualized in Figure \ref{QinQoutWnet_Stir}(b). We can see that for each $q$ value computed, $Q_{out}/k_{B} T_{c}$ value decreases linearly as $T_{h}$ increases, but this trend abruptly changes as $T_{h}$ further increases and the sign of the $Q_{out}/k_{B} T_{c}$ changes from positive to negative. For negative values of $Q_{out}/k_{B} T_{c}$ the decrease as $T_{h}$ decreases is not linear, showing a asymptotic behavior. The behavior is more prominent for the case $q=1$, i.e. the harmonic oscillator, for which the values $Q_{out}/k_{B} T_{c}$ show a negligible dependence on the hot bath temperature $T_{h}$.

Finally, the reduced net Work per cycle $W_{Stir}/k_{B} T_{c}$ is displayed in Figure \ref{QinQoutWnet_Stir}(c), from which is straightforward to see that it shows a linear increase as $T_{h}$ increases for all the $q$ values computed. The role played the $q$ parameter is to regulate the sensitivity of $W_{Stir}/k_{B} T_{c}$ with $T_{h}$ parameter; i.e.  as the $q$ parameter increases, the rate of increase for the $W_{Stir}/k_{B} T_{c}$ also increases.

\begin{figure*}
		\centering
		\includegraphics[scale=0.5]{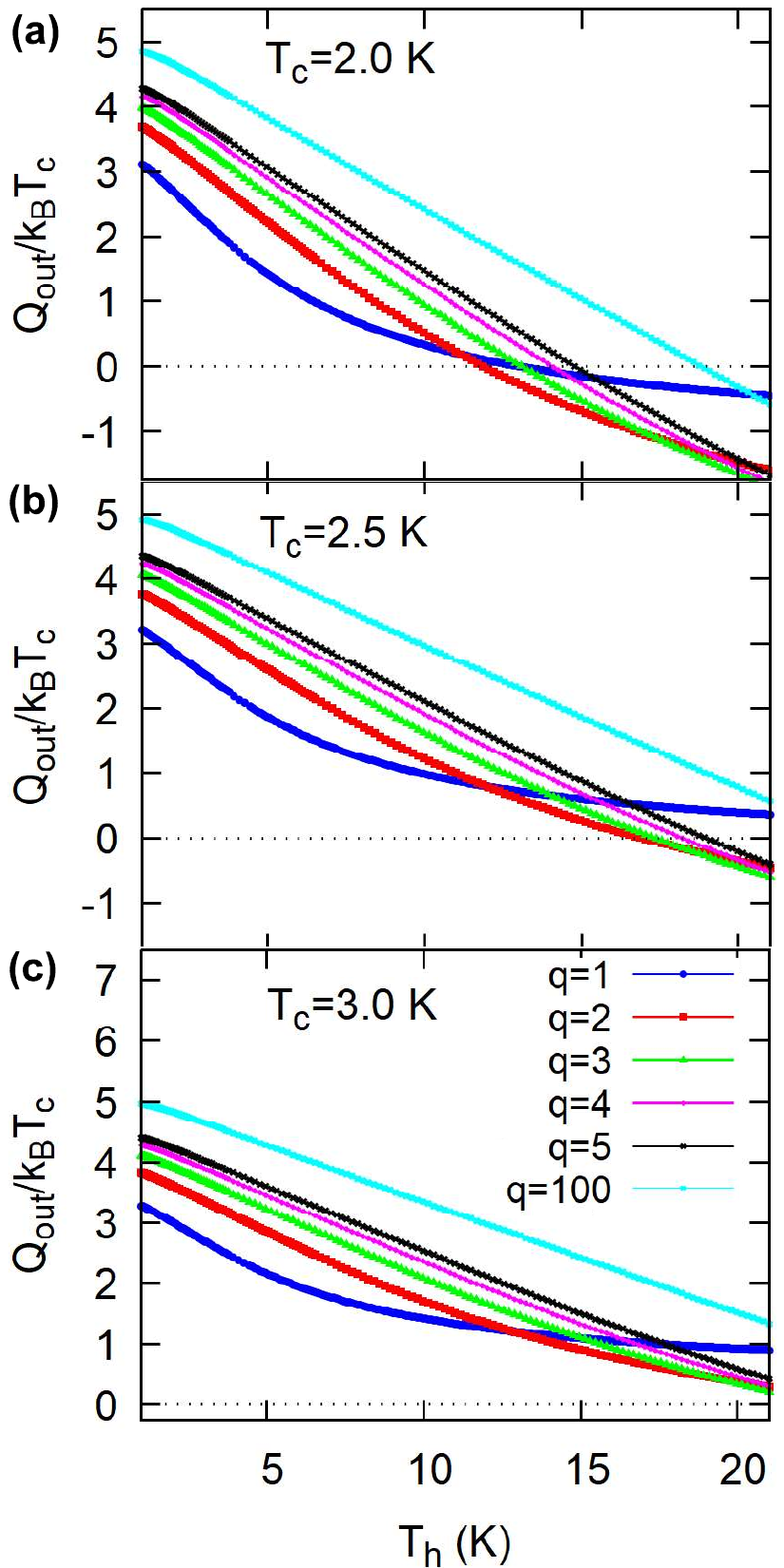}
		\caption{Dependence of $Q_{out}/k_{B} T_{c}$ as a function of $T_{h}$ for distinct $q$ values fixing the cold bath at (a) $T_{c}=2.0$ K, (b) $T_{c}=2.5$ K and (c) $T_{c}=3.0$ K, for the quantum Stirling cycle of the $V(x)=V_0\left(\frac{x}{2a}\right)^{2q}$. Different colors correspond to various $q$ values, each one representing a specific power-law potential. }
    \label{Qout_Tc_Stir}
\end{figure*}

The role played by the temperature of the cold bath $T_{c}$ on the behavior of $Q_{in}/k_{B} T_{c}$, $Q_{out}/k_{B} T_{c}$, and $W_{Stir}/k_{B} T_{c}$ was investigated by varying the values of $T_{c}$. From a qualitative point of view, the behavior dependence of of $Q_{in}/k_{B} T_{c}$ and $W_{Stir}/k_{B} T_{c}$ with $T_{h}$ and $q$ values are unaffected. However the inversion of the heat flux $Q_{out}/k_{B} T_{c}$ deserves a closer look. From Figure \ref{Qout_Tc_Stir}(a), (b), and (c), we see that the $T_{h}$ value required for its sign to change occurs for each $q$ value is shifted to higher values.

\begin{figure*}
		\centering
		\includegraphics[scale=0.55]{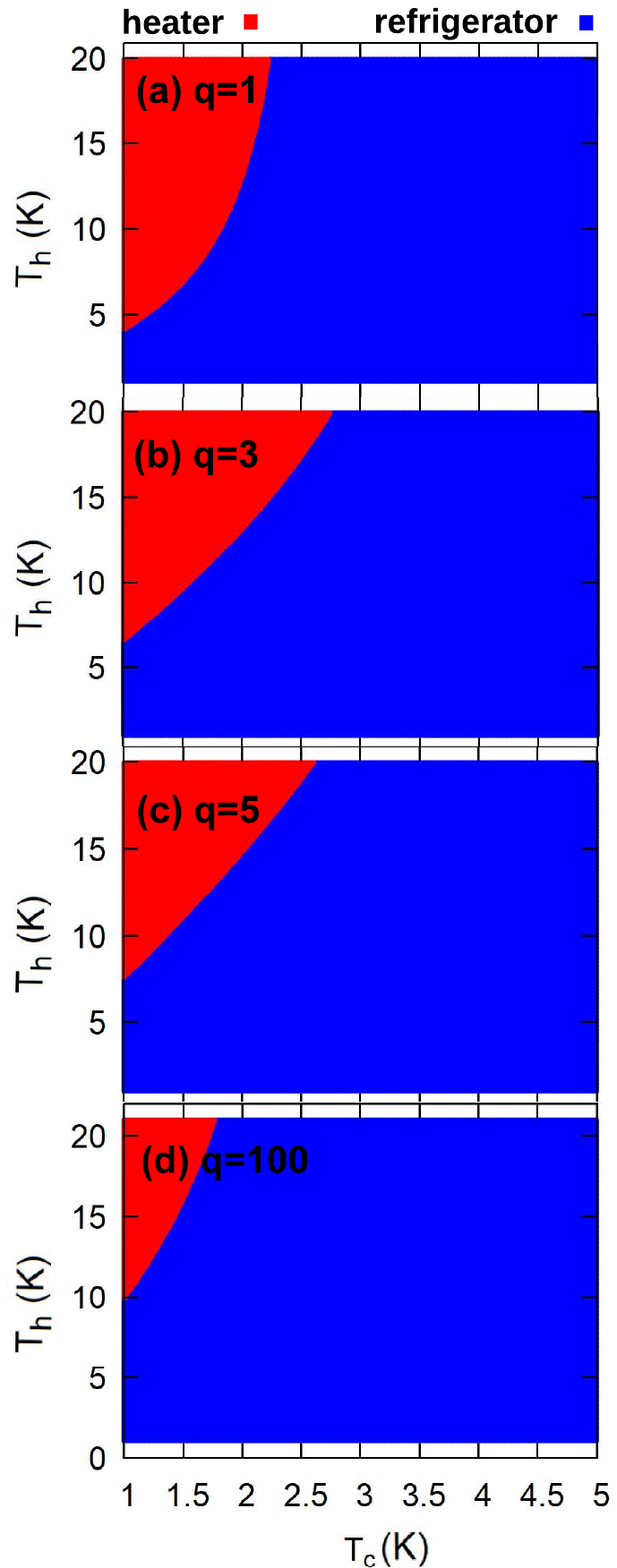}
		\caption{Operation modes obtained for the quantum Stirling cycle of the $V(x)=V_0\left(\frac{x}{2a}\right)^{2q}$ potential, mapping the values of cold 
 and hot bath temperatures $T_{h}$ and $T_{c}$, respectively. The values of the $q$ parameter are (a) $q=1$, (b) $q=3$, (c) $q=5$, and (d) $q=100$. }
    \label{Map_Stir}
\end{figure*}

Changes in sign in one or more of the reduced thermodynamic quantities $Q_{in}/k_{B} T_{c}$, $Q_{out}/k_{B} T_{c}$, and $W_{Stir}/k_{B} T_{c}$ as we tune the values of $T_{h}$, $T_{c}$, and $q$ reveal the occurrence of inversion of the heat fluxes in the system, resulting in different thermodynamic operation regimes that can be leveraged for practical applications. For our quantum Stirling machine, the change in the sign of $Q_{out}/k_{B} T_{c}$ shown in Figures \ref{QinQoutWnet_Stir} and \ref{Qout_Tc_Stir} clearly evidences the presence of at least two operation modes. In this regard, we mapped the sign of $Q_{in}/k_{B} T_{c}$, $Q_{out}/k_{B} T_{c}$, and $W_{Stir}/k_{B} T_{c}$ changing the values of $T_{c}$ and $T_{h}$ values while keeping the $q$ parameter fixed. This approach led to the results displayed in Figures \ref{Map_Stir}(a), (b), , and (d) which reveals that upon proper selection of the $T_{c}$ and $T_{h}$ values it is possible to set the system to operate either as a refrigerator or as a heater.

The refrigerator mode is characterized by work done \textit{on} ($\Delta W<0$) the working medium, while it absorbs the heat from the cold bath ($Q_{in}>0$) releasing it in the hot bath ($Q_{in}<0$). On the heater mode work is also performed \textit{ on} ($\Delta W<0$), but now the working medium releases heat to both thermal reservoirs (therefore $Q_{in}<0$ and $Q_{out}<0$). Finally, small region on the $T_{h} - T{c}$ diagram corresponding to the heater mode changes slightly as the $q$ value increases. We conclude that the operation modes are predominantly defined by tuning the thermal baths regardless of the $q$ parameter. Finally, in Figure \ref{Map_Stir} (d) we see that as we increase the value of $q$, the refrigerator mode is favored, since the region of the $T_{h} - T{c}$ diagram to corresponding the heater mode diminishes.

The efficiency of the quantum Stirling machine is computed through the following expression:
\begin{align} \label{eta_Stirling}
    \eta = \frac{\frac{T_{h}}{T_{c}} \left[\ln (2) + \frac{4q}{q+1} \right] + \ln (2) + \frac{4q}{q+1}}{-\frac{T_{h}}{T_{c}} \left[\ln(2) + \frac{4q}{q+1} \right] +  \frac{ C_q}{2 k_{B} T_{c}} \sum_{n=1}^{\infty} n^{\frac{2q}{q+1}}\left\{\frac{\exp\left(-\frac{ C_qn^{\frac{2q}{q+1}}}{k_B T_h}\right)}{ \Theta_{h} } + \frac{ \exp\left(-\frac{ C_qn^{\frac{2q}{q+1}}}{k_B T_c}\right)}{ \Theta_{c}} \right\}} .
\end{align}

\begin{figure*}
		\centering
		\includegraphics[scale=0.5]{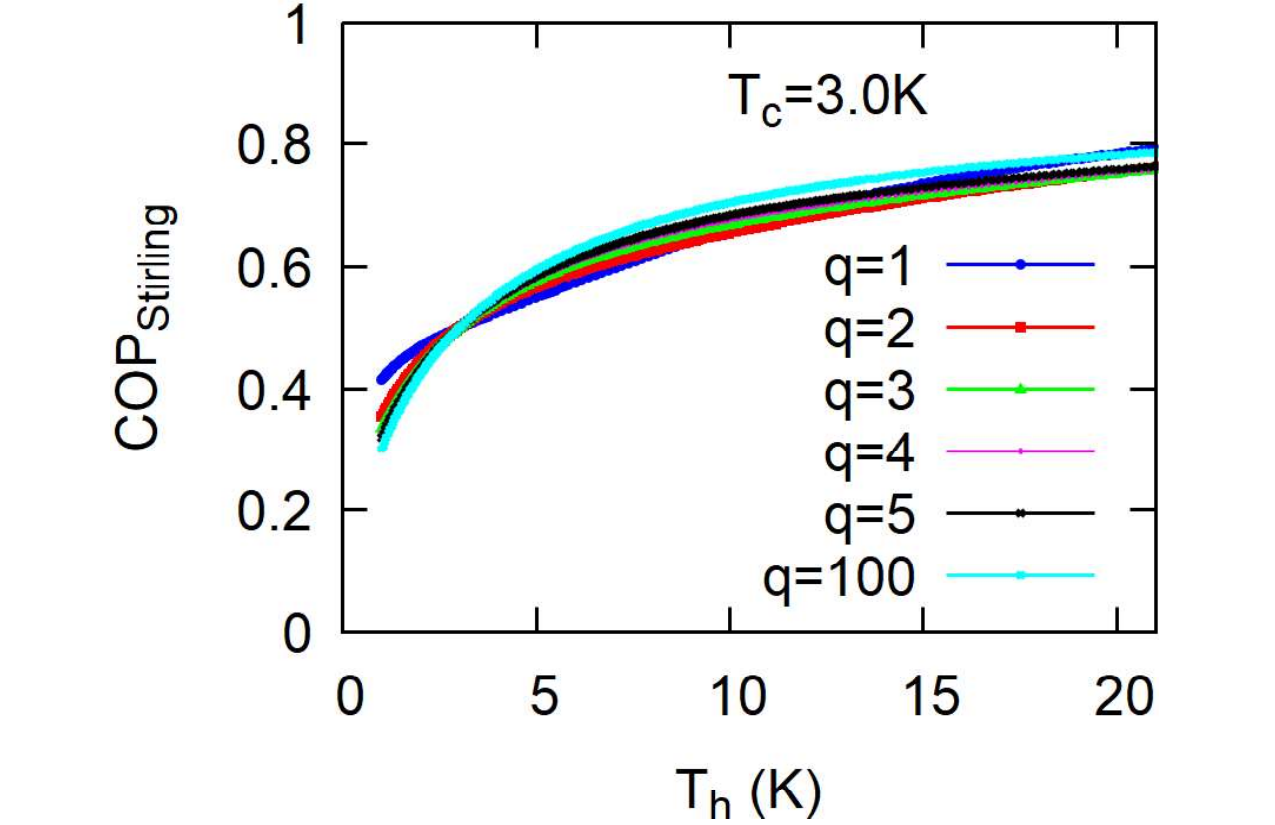}
		\caption{COP values for the quantum Stirling cycle of the $V(x)=V_0\left(\frac{x}{2a}\right)^{2q}$ potential, plotted as a function of the $T_{h}$ for different $q$ values, for a fixed cold bath temperature of $T_{c}=3.0$ K. }
    \label{COP_Stir}
\end{figure*}

Due to the presence of the refrigerator mode, we calculated the coefficient of performance \textit{COP}, defined by the following expression:

\begin{align} \label{COP_Stirling}
    COP = \frac{\Delta Q_{in}}{W_{Stir}}.
\end{align}
The values of $COP$ as a function of the hot bath temperature $T_{h}$ are shown in Figure \ref{COP_Stir} for a fixed cold bath temperature of $T_{c}=3.0$ K and in a hot bath temperature interval $1.0 K \leq T_{h} \leq 20.0$ K. For this entire range of values, the system operates as a refrigerator, regardless of the $q$ values (see Figure \ref{Map_Stir}). 

\section{Quantum Otto Cycle}
Changing the isothermal processes of a Stirling cycle with adiabatic ones, we are left with a quantum Otto cycle, combining two quantum adiabatic and two isochoric processes. For this cycle, there is only heat exchange and work done for each stroke, opening the possibility of understanding the nature of both quantities independently. From a mathematical point of view, it is easy to model the quantum Otto cycle by simply calculating the heat fluxes $Q_{in}$ and $Q_{out}$ in addition to the net work per cycle $W_{Stir}$ and the corresponding efficiency $\eta_{Otto}$. The modelling of Otto Quantum thermal machines have already been explored in the literature \cite{sukamto2023effect,rodin2024three}.  We obtain the corresponding thermodynamic expression in terms of occupation probabilities $p_n$ and energy spectra $E_n$ given by equations \ref{pn_thermal_states}  and \ref{En}, respectively, following reference \cite{Quan:07}. In this way, we can model the Otto cycle in terms of thermalized states with the hot or cold reservoir (for details in the calculation, see the appendix \ref{AP:Otto_limits}). In this approach to model the Otto cycle, in addition to $T_{c}$, $T_{h}$, and $q$, an additional parameter $r$ arises:

\begin{align} \label{r_def}
      r = \frac{\omega_{h}}{\omega_{c}}
\end{align}
where $\omega_{h}$ ($\omega_{c}$) stand for the respective frequency under thermal equilibrium with the hot (cold) bath. It can be obtained by simply calculating the ratio between the energy spectra (given by Equation \ref{En}) considering both thermal equilibrium cases.

For the reduced heat fluxes $Q_{in}$ and $Q_{out}$ for the $V(x)=V_0\left(\frac{x}{2a}\right)^{2q}$ potential we have the following expressions \cite{quan2009quantum, Quan:07}:
\begin{align} \label{Qin_Otto}
      Q_{in}(q,T_h, T_c,r) = \frac{C_q}{k_{B} T_{c}} \sum_{n} \left\{\left|n\right|^{\frac{2q}{q+1}}  
\left[ 1- \exp\left(\frac{C_q n^{\frac{2q}{q+1}}  }{k_{B} T_{h}}-\frac{C_q\left|n\right|^{\frac{2q}{q+1}}}{k_{B} T_{c} r^{\frac{2q}{q+1}}}\right) \right] \right\}
\end{align}
 and
\begin{align} \label{Qout_Otto}
            Q_{out}(q,T_h, T_c,r) = \frac{C_q}{k_{B} T_{c} r^{\frac{2q}{q+1}}} \sum_{n} \left\{n^{\frac{2q}{q+1}}  
\left[\exp\left(\frac{C_q\left|n\right|^{\frac{2q}{q+1}}  }{k_{B} T_{h}}-\frac{C_q\left|n\right|^{\frac{2q}{q+1}}}{k_{B} T_{c} r^{\frac{2q}{q+1}}}\right) - 1 \right] \right\}
\end{align}
where $E_n^{h}$($p_n^{h}$) refer to the energy level (occupation probability) in thermal equilibrium with the hot bath $T_{c}$, and $E_n^{c}$($p_n^{c}$) are the same quantities but now considering thermal equilibrium with the cold bath. Following this approach, the net work per cycle is given by:

\begin{align} \label{Wnet_Otto}
      \Delta W_{Otto}(q,T_h, T_c,r)=\left(\frac{1}{r^{\frac{2q}{q+1}}}-1\right)\frac{ C_q}{k_{B} T_{c}} \sum_{n} \left\{\left|n\right|^{\frac{2q}{q+1}}  
\left[\exp\left(\frac{C_q\left|n\right|^{\frac{2q}{q+1}}  }{k_{B} T_{h}}-\frac{C_q\left|n\right|^{\frac{2q}{q+1}}}{k_{B} T_{c} r^{\frac{2q}{q+1}}}\right) -1\right] \right\}
\end{align}

for which detailed calculations can be seen in the Appendix \ref{AP:Otto_limits}.

\begin{figure*}
		\centering
		\includegraphics[scale=0.5]{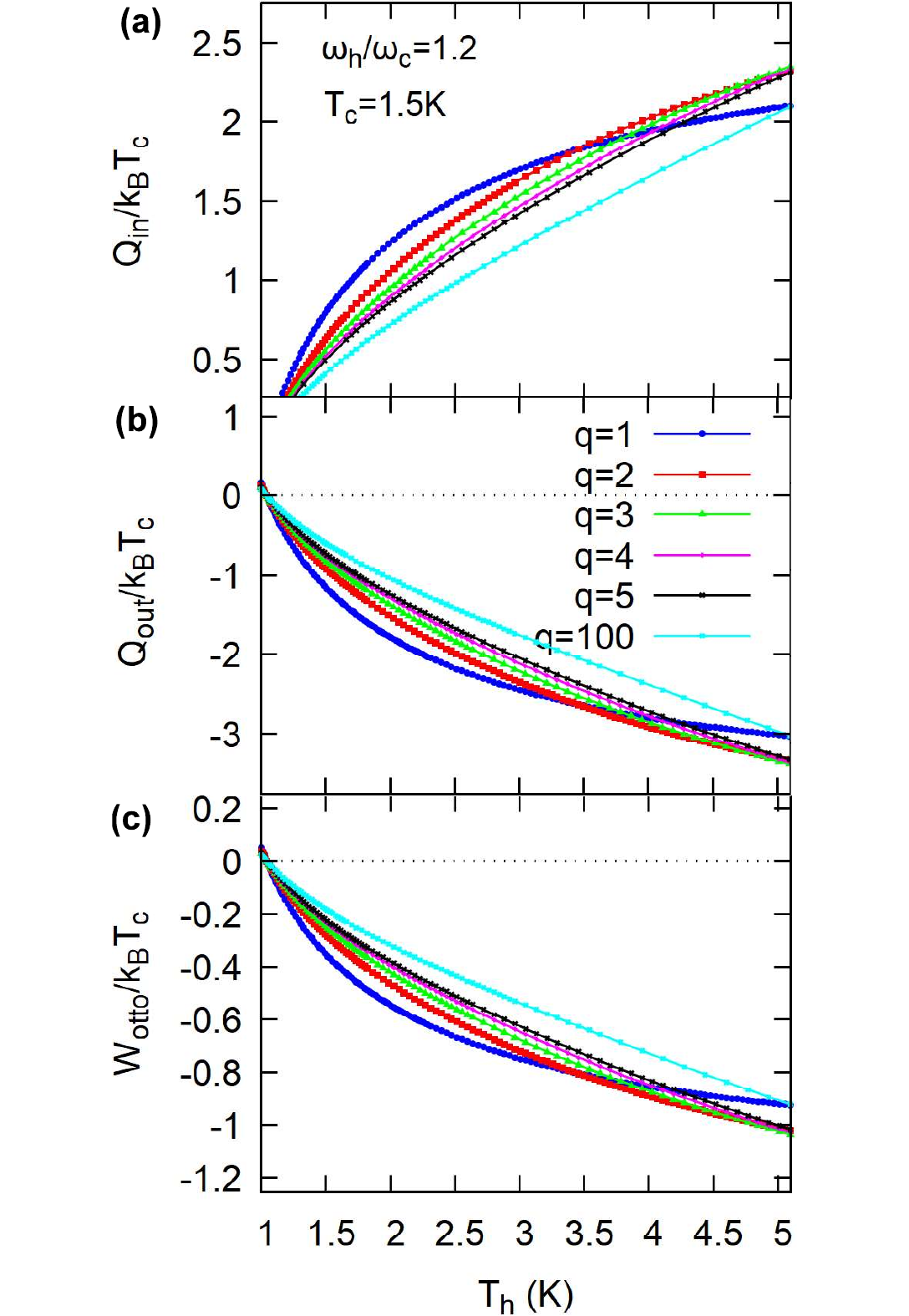}
		\caption{Thermodynamic quantities of interest (a) $Q_{in}/k_{B} T_{c}$, (b) $Q_{out}/k_{B} T_{c}$, and (c) $W_{net}/k_{B} T_{c}$, for the quantum Otto cycle of the $V(x)=V_0\left(\frac{x}{2a}\right)^{2q}$ potential, plotted as a function of the $T_{h}$ for a fixed cold bath temperature of $T_{c}=1.5$ K. The frequency ratio was fixed at $\omega_{h}/\omega_{c}=1.2$ for obtan the curves displayed here.}
    \label{QinQoutWnet_Otto}
\end{figure*}

\begin{figure*}
		\centering
		\includegraphics[scale=0.5]{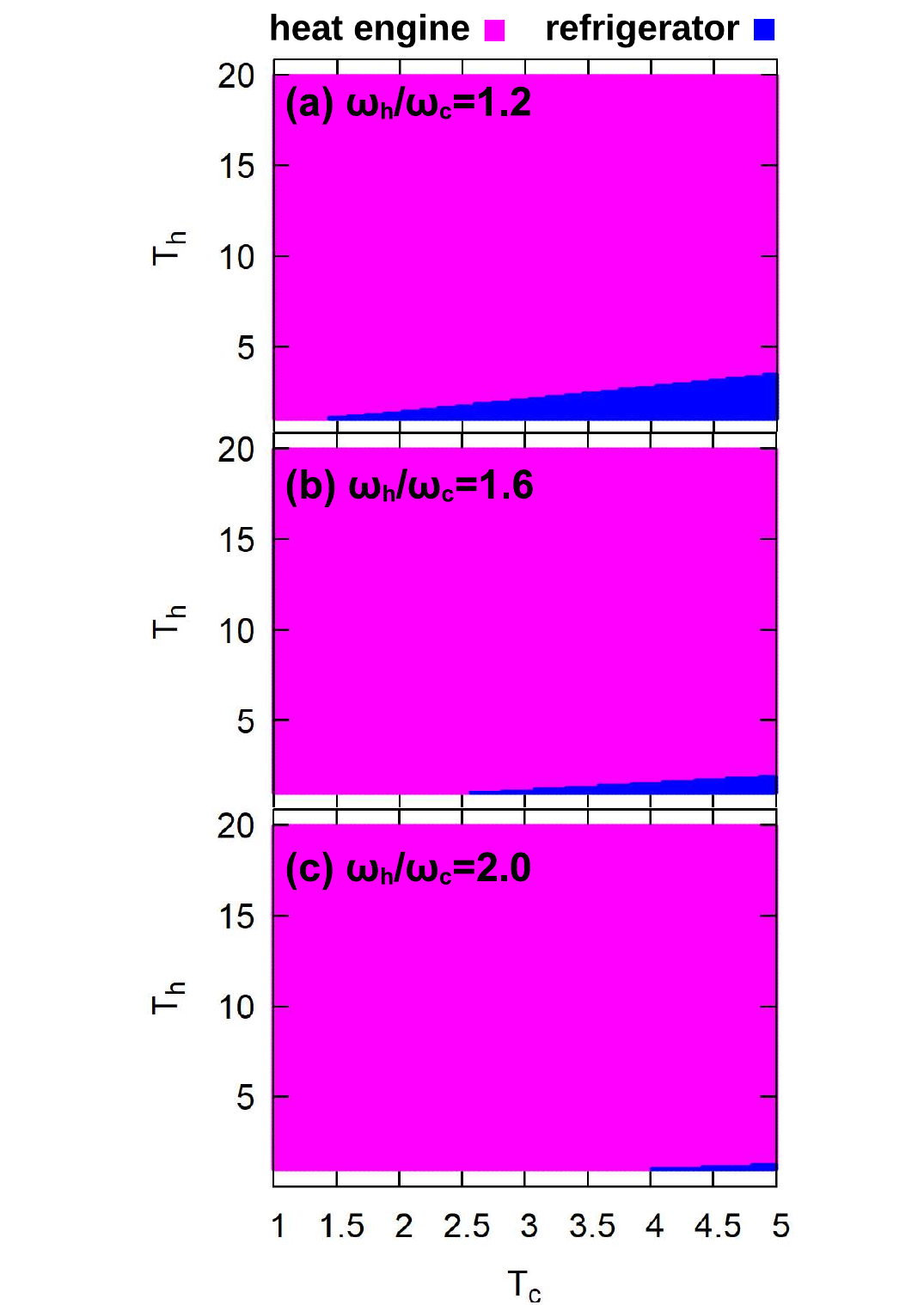}
		\caption{Operation modes obtained for the quantum Otto cycle of the $V(x)=V_0\left(\frac{x}{2a}\right)^{2q}$ potential, for a fixed value of $q=3.0$. The values of the $\omega_{h}/\omega_{c}$ ratio are: (a) $\omega_{h}/\omega_{c}=1.2$, (b) $\omega_{h}/\omega_{c}=1.6$, and (c) $\omega_{h}/\omega_{c}=2.0$. }
    \label{Map_Otto_Tc1}
\end{figure*}

To investigate the role of these parameters in the reduced thermodynamic quantities, we plotted the values of $Q_{in}/k_{B} T_{c}$, $Q_{out}/k_{B} T_{c}$, and $W_{Otto}/k_{B} T_{c}$, as a function of $T_{h}$ fixing the values of $T_{c}=1.5$ K and $\omega_{h}/\omega_{c}=1.2$, as shown in Figure \ref{QinQoutWnet_Otto} (a), (b) and (c), respectively. A curious trend observed is that the three quantities change sign at the same $T_{h}$ value, suggesting the presence of different operating regimes.
We have mapped the system in search for distinct operation modes also in the Otto quantum cycle, varying the values of all four relevant parameters: $T_{c}$, $T_{h}$, $\omega_{h}/\omega_{c}$, and $q$ values. Figure \ref{Map_Otto_Tc1} shows the results obtained fixing the temperature of the cold bath at $T_{c}=1.5$ K and $T_{c}=3.0$ K. We can observe a transition between two distinct modes: heater and refrigerator.

In addition, the role of $T_{c}$ in the operation modes is evidenced; we noted that it drastically impacts the transition region between the refrigerator and the heat engine modes. Increasing the values of $T_{c}$K expands the region of $T_h$-$T_{c}$ diagram in which the system operates as a refrigerator. In other words, the value of $T_{c}$ acts as an upper temperature limit for the refrigerator mode. The physical interpretation is that for regions where the thermal baths are inverted $T_{c}>T_{h}$ only the refrigerator regime is possible, while when the condition $T_{c}<T_{h}$ holds, it is possible to induce the heat engine mode with proper selection of the values $r$ and $T_{h}$.

\begin{figure*}
		\centering
		\includegraphics[scale=0.5]{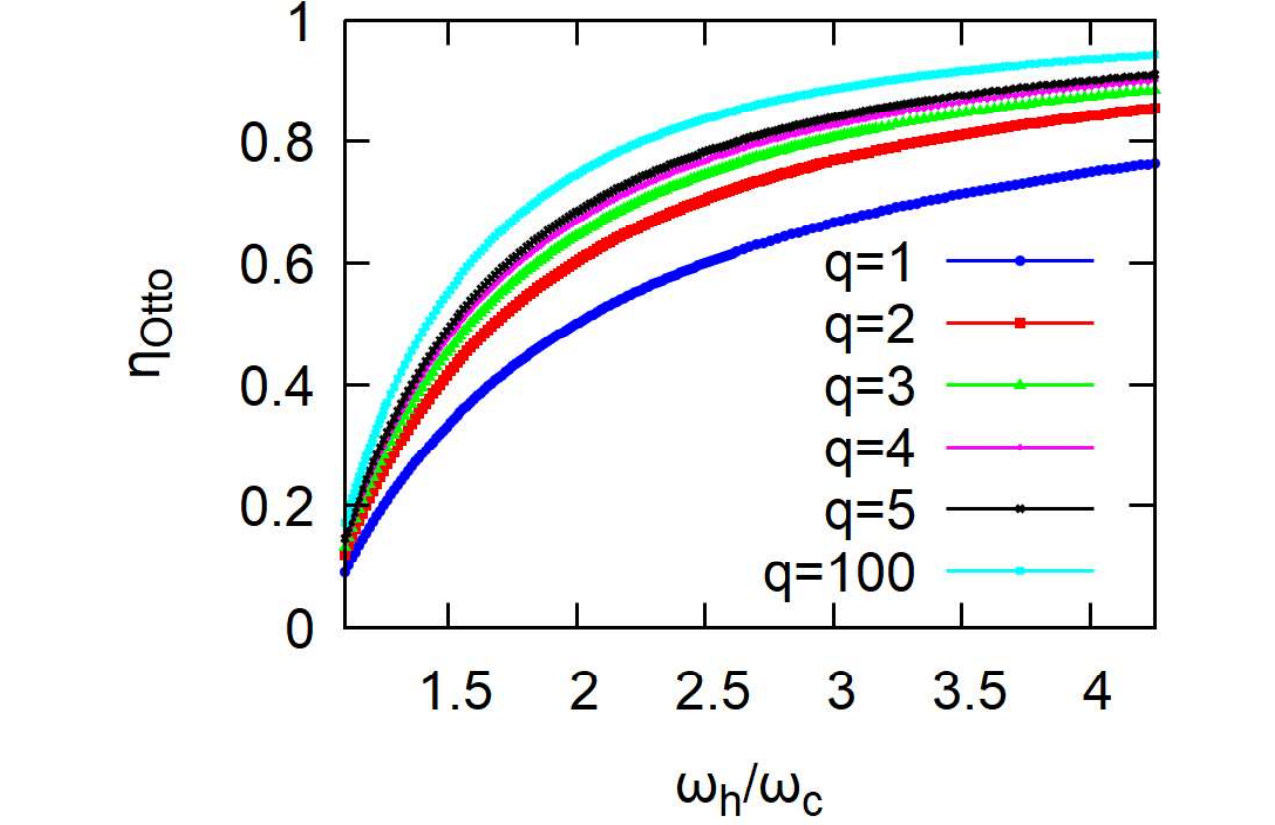}
		\caption{ $\eta$ values for the quantum Otto cycle of the $V(x)=V_0\left(\frac{x}{2a}\right)^{2q}$ potential, plotted as a function of the $\omega_{h}/\omega_{c}$ ratio for different $q$ values. }
    \label{eta_Otto}
\end{figure*}

Finally, the efficiency of our Otto cycle is given by:
\begin{align} \label{eff_Otto}
      \eta_{Otto}(q,r)=\frac{\left(r^{\frac{q}{q+1}}-1  \right)\left(r^{\frac{q}{q+1}}+1  \right)}{r^{\frac{2q}{q+1}}}
\end{align}
and is plotted as a function of $\omega_{h}/\omega_{c}$ ratio in figure \ref{eta_Otto}. The non-dependency of $\eta_{Otto}(q,r)$ with the temperature of the thermal baths is in accordance with the literature \cite{Quan:07}, since it is expected to depend only on the ratio between the energies $E_{n}^{h}$ and $E_{n}^{c}$, which in this case is the $\omega_{h}/\omega_{c}$ ratio.

\newpage

\section{Conclusions and Perspectives}

In this work, we performed numerical modeling of Stirling and Otto quantum machines using the generic power-like potential $V(x)=V_0\left(\frac{x}{2a}\right)^{2q}$ as a working substance. A general trend observed for both Stirling and Otto cycles is that the dependence of thermodynamic properties on the $q$ values continues up to $q=3$. The $q$-dependent energy spectra correctly recover the energies for the 1-D harmonic oscillator and infinite potential well cases. Mapping the behavior of $Q_{in}/k_{B} T_{c}$, $Q_{out}/k_{B} T_{c}$, and $W_{net}/k_{B} T_{c}$ as a function of the hot $T_{h}$ and cold $T_{c}$ baths and the $q$ parameter allows us to show that the Stirling machine can operate between heater and refrigerator modes. The occurrence of these two modes is consequence of the inversion in the heat flux between the cold bath and the working substance $Q_{out}/k_{B} T_{c}$, largely due to changes in the thermal baths, regardless of the $q$ value. A similar analysis for the Otto machine required an additional investigation of the role played by the $\omega_h/\omega_c$ ratio, which is the work parameter for our Otto cycle. In this case, the system can operate in the heat engine or refrigerator modes, where the heat engine mode is favored by increasing the values $\omega_h/\omega_c$. An interesting perspective is to discuss possible physical interpretations of the cases $q=2$ and $q=3$, as their thermodynamic behavior deviates from each other.

\begin{acknowledgments}
V.G de Paula thanks the Rio de Janeiro State Research Support Foundation (FAPERJ) for financial support.  C. Cruz thanks the Bahia State Research Support Foundation (FAPESB) for financial support. This study was financed in part by the Coordenação de Aperfeiçoamento de Pessoal de Nível Superior – Brasil (CAPES) – Finance Code 001. MSR thanks FAPERJ and CNPq for financial support.  
\end{acknowledgments}

\newpage

\appendix

\section{Appendix Stirling cycle model} \label{AP:Stirling}

Due to the symmetric nature of our generic power-like potential, when imposing the quantum isothermal expansion (i.e. inserting the barrier in the middle point of the potential), it results in a new partition function that is multiple of the older one. Let's consider a particle trapped in a box (1-D potential well) of length $2a$, we have:

\begin{align}
E_n=n^2 \frac{\hbar^2 \omega}{2m(2a)^2}
\end{align}
which leads to the following partition function:
\begin{align}
Z=\sum_{n=1}^{\infty} \exp\left({-\frac{E_n}{k_B T_h}}\right)=\exp\left({-\frac{n^2 \frac{\hbar^2 \omega}{2m(2a)^2}}{k_B T_h}}\right)
\end{align}

After inserting the barrier, we now have two boxes, each one with length $a$. The even energy levels arise to match the next odd energy levels, which means each odd energy level is now double degenerated:
\begin{align}
 E_n'=n'^2 \frac{\hbar^2 \omega}{2m(2a)^2}=E_n=(2n)^2 \frac{\hbar^2 \omega}{2m(2a)^2} 
\end{align}

The new partition function after the quantum isothermal expansion:

\begin{align}
 Z'=\sum_{n=1}^{\infty} 2\exp\left({-\frac{E_n'}{k_B T_h}}\right)=2\exp\left({-\frac{n^2 \frac{\hbar^2 \omega}{2m(2a)^2}}{k_B T_h}}\right)
\end{align}

Extending this to our generic power-like potential (which displays similar energy spectra) we obtain the partition functions $Z_B$ and $Z_C$ given by equations \ref{Z_B_Stirling} and \ref{Z_C_Stirling}, respectively. Complementarily,  the partition functions $Z_A$
and $Z_B$ given by equations \ref{Z_A_Stirling} and \ref{Z_B_Stirling} are obtained combining equation \ref{Z_thermal_states}, \ref{Z_h} and \ref{Z_c}.
The heat exchanges for the A $\rightarrow B$ isothermal process are given as follows:
\begin{align} \label{AP:Q_AB_Stirling}
   \frac{\Delta Q_{AB}(q,T_h,T_c)}{k_{B} T_{c}} =\left( \frac{1}{k_{B} T_{c}}\right) \left[ \underbrace {\frac{\partial{\left(\ln{\frac{Z_A}{Z_B}}\right)}}{\partial{\beta_h}}}_{\Delta U_{A,B}} + \underbrace {k_B T_h \ln{\left(\frac{Z_A}{Z_B}\right)}}_{W_{A,B}} \right] = -\frac{T_{h}}{T_{c}} \left\{\ln(2) + \frac{4q}{q+1} \right\}
\end{align}
For the B$\rightarrow$C isochoric process, the First Law gives us:
\begin{align} \label{AP:Q_BC_Stirling}
   \frac{\Delta Q_{BC}(q,T_h,T_c)}{k_{B} T_{c}} = \frac{\Delta U_{B,C}}{k_{B} T_{c}} = \left( \frac{1}{k_{B} T_{c}} \right) \left[ -\frac{\partial{(\ln{Z_C})}}{\partial{\beta_c}} + \frac{\partial{(\ln{Z_B})}}{\partial{\beta_h}} \right] = \frac{ C_q}{2 k_{B} T_{c}} \sum_{n=1}^{\infty} n^{\frac{2q}{q+1}}\left\{\frac{\exp\left(-\frac{ C_qn^{\frac{2q}{q+1}}}{k_B T_c}\right)}{ \Theta_c } + \frac{ \exp\left(-\frac{ C_qn^{\frac{2q}{q+1}}}{k_B T_h}\right)}{ \Theta_h} \right\}
\end{align}

For the isothermal stroke $C \rightarrow D$ we have:

\begin{align} \label{AP:Q_CD_Stirling}
  \frac{\Delta Q_{CD}(q,T_h,T_c)}{k_{B} T_{c}} =\left( \frac{1}{k_{B} T_{c}}\right) \left\{ \underbrace {\frac{\partial{\left(\ln{\frac{Z_D}{Z_C}}\right)}}{\partial{\beta_h}}}_{\Delta U_{D,C}} + \underbrace {k_B T_h \ln{\left(\frac{Z_D}{Z_C}\right)}}_{W_{C,D}}\right\}  = -\frac{4q}{q+1}
\end{align}

Finally, the last isochoric stroke D $\rightarrow$ A gives us the following expressions:

\begin{align} \label{AP:Q_DA_Stirling}
    \Delta Q_{DA}(q,T_h,T_c)= \Delta U_{D,A} = -\frac{\partial{(\ln{Z_A})}}{\partial{\beta_c}} + \frac{\partial{(\ln{Z_D})}}{\partial{\beta_h}} = \frac{ C_q}{2 k_{B} T_{c}} \sum_{n=1}^{\infty} n^{\frac{2q}{q+1}}\left\{\frac{\exp\left(-\frac{ C_qn^{\frac{2q}{q+1}}}{k_B T_h}\right)}{ \Theta_{h} } + \frac{ \exp\left(-\frac{ C_qn^{\frac{2q}{q+1}}}{k_B T_c}\right)}{ \Theta_{c}} \right\}
\end{align}

For a suitable analysis of the thermodynamic properties of the system, we have also calculated the reduced heat fluxes of $Q_{in}/k_{B} T_{c}$ and $Q_{out}/k_{B} T_{c}$, which account for the total heat changed with the hot and cold reservoirs, respectively:

\begin{align} \label{AP:Q_in_Stirling}
   \Delta Q_{in}(q,T_h) = \frac{\Delta Q_{AB}(q,T_h,T_c)}{k_{B} T_{c}}+\frac{\Delta Q_{DA}(q,T_h,T_c)}{k_{B} T_{c}} = \frac{T_{h}}{T_{c}} \left[\ln(2) + \frac{4q}{q+1} \right] +  \frac{C_q}{2 k_{B} T_{c}} \sum_{n=1}^{\infty} n^{\frac{2q}{q+1}}\left(\frac{\exp\left(-\frac{ C_qn^{\frac{2q}{q+1}}}{k_B T_h}\right)}{ \Theta_h } + \frac{ \exp\left(-\frac{ C_qn^{\frac{2q}{q+1}}}{k_B T_c}\right)}{ \Theta_c} \right)
\end{align}

The reduced heat fluxes are easily obtained from the previous expressions:

\begin{align} \label{AP:Q_out_Stirling}
    \Delta Q_{out}(q,T_c) = \frac{\Delta Q_{BC}(q,T_h,T_c)}{k_{B} T_{c}}+\frac{\Delta Q_{DA}(q,T_h,T_c)}{k_{B} T_{c}} =\frac{ C_q}{2 k_{B} T_{c}} \sum_{n=1}^{\infty} n^{\frac{2q}{q+1}}\left\{\frac{\exp\left(-\frac{ C_qn^{\frac{2q}{q+1}}}{k_B T_c}\right)}{ \Theta_{c} } + \frac{ \exp\left(-\frac{ C_qn^{\frac{2q}{q+1}}}{k_B T_h}\right)}{ \Theta_{h}} \right\} -\frac{4q}{q+1}
\end{align}

The net work done per cycle is easily calculated by adding the corresponding terms in Equations \ref{Q_AB_Stirling} and \ref{Q_CD_Stirling}:

\begin{align} \label{AP:Wnet_Stirling}
       \frac{\Delta W_{Stir}}{k_{B} T_{c}} =\frac{ W_{AB} + W_{CD}}{k_{B} T_{c}} = \frac{T{_h}}{T{_c}}  \ln{\left( \frac {Z_B}{Z_A} \right) }  -\ln{\left( \frac {Z_D}{Z_C}\right)}=\frac{T{_h}}{T{_c}}  \ln{\left[ 2 \exp(\frac{4q}{q+1}) \right] } - \ln{ \left[ \frac{\exp(-\frac{4q}{q+1})}{2} \right]} = \frac{T{_h}}{T{_c}}\ln(2) +\frac{T{_h}}{T{_c}} \frac{4q}{q+1} + \frac{4q}{q+1} + \ln(2) =\left(\frac{T_{h}}{T_{c}}+1\right) \left(ln(2) + \frac{4q}{q+1} \right)
\end{align}

It is important to note that imposing the high-energy/low-temperature limit ($E_{n}/k_{B}T >> 1$) on the equation \ref{eta_Stirling} we recover the Carnot limit for the quantum Stirling cycle:

\begin{align} \label{AP:eta_stir}
    \eta &= \frac{\frac{T_{h}}{T_{c}} \left[\ln (2) + \frac{4q}{q+1} \right] + \ln (2) + \frac{4q}{q+1}}{-\frac{T_{h}}{T_{c}} \left(\ln(2) + \frac{4q}{q+1} \right) + \underbrace{\frac{ C_q}{2 k_{B} T_{c}} \sum_{n=1}^{\infty} n^{\frac{2q}{q+1}}\left\{\frac{\exp\left(-\frac{ C_qn^{\frac{2q}{q+1}}}{k_B T_h}\right)}{ \Theta_{h}} + \frac{ \exp\left(-\frac{ C_qn^{\frac{2q}{q+1}}}{k_B T_c}\right)}{ \Theta_{c}} \right\}}}_{\rightarrow 0}  \\
    &\approx  1+\left\{ \frac{\frac{-4q}{q+1}-ln(2)}{\frac{T_{h}}{T_{c}} \left[ ln(2)+\frac{-4q}{q+1}\right]}\right\} \approx 1+\frac{T_{c}}{T_{h}} \left(\frac{\frac{-4q}{q+1}-ln(2)}{\frac{-4q}{q+1}+ln(2)} \right) \approx  1-\frac{T_{c}}{T_{h}} \left(\frac{\frac{4q}{q+1}-ln(2)}{\frac{4q}{q+1}-ln(2)} \right) \approx 1 - \left(\frac{T_{c}}{T_{h}} \right)
\end{align}

\section{Appendix Otto cycle model} \label{AP:Otto_limits}

For the Otto cycle, the reduced heat changed with the hot bath is : 
\begin{align} \label{AP:Qin_Otto}
      Q_{in} = \frac{1}{k_{B} T_{c}} \sum_{n} E_n^{h}(p_n^{h}-p_n^{c}) = \frac{C_q}{k_{B} T_{c}} \sum_{n} \left\{\left|n\right|^{\frac{2q}{q+1}}  
\left[ 1- \exp\left(\frac{C_q\left|n\right|^{\frac{2q}{q+1}}  }{k_{B} T_{h}}-\frac{C_q\left|n\right|^{\frac{2q}{q+1}}}{k_{B} T_{c} r^{\frac{2q}{q+1}}}\right) \right] \right\}
\end{align}

In a similar fashion, the reduced heat changed with the cold reservoir is:

\begin{align} \label{AP:Qout_Otto}
      Q_{out} = \frac{1}{k_{B} T_{c}} \sum_{n} E_n^{c}(p_n^{c}-p_n^{h}) = \frac{C_q}{k_{B} T_{c}} \sum_{n} \left\{\left|n\right|^{\frac{2q}{q+1}}  
\left[\exp\left(\frac{C_q\left|n\right|^{\frac{2q}{q+1}}  }{k_{B} T_{h}}-\frac{C_q\left|n\right|^{\frac{2q}{q+1}}}{k_{B} T_{c} r^{\frac{2q}{q+1}}}\right) - 1 \right] \right\}
\end{align}

The reduced net work per cycle is given simply by summing up $Q_{in}$ and $Q_{out}$, and noting that $Q_{in}$ and $Q_{out}$ must have opposite signs:

\begin{align} \label{AP:Wnet_Otto}
    \frac{ \Delta W_{Otto}}{k_{B} T_{c}} =\frac{Q_{in} - Q_{out}}{k_{B} T_{c}}=\frac{1}{k_{B} T_{c}} \sum_{n} (E_n^{h}-E_n^{c})(p_n^{h}-p_n^{c}) =\left(\frac{1}{r^{\frac{2q}{q+1}}}-1\right)\frac{ C_q}{k_{B} T_{c}} \sum_{n} \left\{\left|n\right|^{\frac{2q}{q+1}}  
\left[\exp\left(\frac{C_q\left|n\right|^{\frac{2q}{q+1}}  }{k_{B} T_{h}}-\frac{C_q\left|n\right|^{\frac{2q}{q+1}}}{k_{B} T_{c} r^{\frac{2q}{q+1}}}\right) -1\right] \right\}
\end{align}

The efficiency of the Otto cycle is calculated as follows:

\begin{align} \label{AP:eff_Otto}
      \eta_{Otto}=\frac{W_{Otto}}{Q_{in}}=\frac{\left(r^{\frac{q}{q+1}}-1  \right)\left(r^{\frac{q}{q+1}}+1  \right)}{r^{\frac{2q}{q+1}}}
\end{align}

and is independent of the hot and cold baths, just as expected.

\end{document}